\documentclass[twocolumn,floatfix]{revtex4}
\usepackage{natbib}
\usepackage{amsmath}
\usepackage{graphicx}
\usepackage{wasysym}

\bibpunct{(}{)}{;}{a}{}{,}

\newcommand{\romn}[1] {{\mathrm #1}}
\newcommand\fs{\hbox{$.\!\!^{\romn s}$}}
\newcommand\farcm{\hbox{$.\mkern-4mu^\prime$}}
\newcommand\farcs{\hbox{$.\!\!^{\prime\prime}$}}

\newcommand{\gr}{$\gamma$-ray}
\newcommand{\grs}{$\gamma$-rays}
\newcommand{\hess}{H.E.S.S.}
\newcommand{\hgcfull}{HESS~J1745-290}
\newcommand{\hgc}{HESS~J1745-290}
\newcommand{\astar}{Sgr~A$^*$}
\newcommand{\aeast}{Sgr~A~East}
\newcommand{\pwn}{G359.95-0.04}
\newcommand{\position}{$\alpha=17^\mathrm{h} 45^\mathrm{m} 39\fs6\pm 0\fs4_\mathrm{stat}\pm 0\fs4_\mathrm{sys}$, 
  $\delta=-29^\circ 0' 22''\pm 6''_\mathrm{stat}\pm 6''_\mathrm{sys}$}
\newcommand{\distToAStar}{$8''\pm 9''_\mathrm{stat}\pm 9''_\mathrm{sys}$}

\begin{document}

\title{Localising the VHE $\gamma$-ray source at the Galactic Centre}

\author{HESS Collaboration}
\author{F.~Acero$^{15}$}
\author{F. Aharonian$^{1,13}$}
\author{A.G.~Akhperjanian$^{2}$}
\author{G.~Anton$^{16}$}
\author{U.~Barres de Almeida$^{8}$} \thanks{supported by CAPES Foundation, Ministry of Education of Brazil}
\author{A.R.~Bazer-Bachi$^{3}$}
\author{Y.~Becherini$^{12}$}
\author{B.~Behera$^{14}$}
\author{K.~Bernl\"ohr$^{1,5}$}
% ...

\vspace{9mm}

\affiliation{
$^1$
Max-Planck-Institut f\"ur Kernphysik, P.O. Box 103980, D 69029
Heidelberg, Germany} \affiliation{
$^2$
 Yerevan Physics Institute, 2 Alikhanian Brothers St., 375036 Yerevan,
Armenia} \affiliation{
$^3$
Centre d'Etude Spatiale des Rayonnements, CNRS/UPS, 9 av. du Colonel Roche, BP
4346, F-31029 Toulouse Cedex 4, France}
% ...
\affiliation{
$^5$
Institut f\"ur Physik, Humboldt-Universit\"at zu Berlin, Newtonstr. 15,
D 12489 Berlin, Germany}
% ...
\affiliation{
$^8$
University of Durham, Department of Physics, South Road, Durham DH1 3LE,
U.K.}
% ...
\affiliation{
$^{12}$
Astroparticule et Cosmologie (APC), CNRS, Universite Paris 7 Denis Diderot,
10, rue Alice Domon et Leonie Duquet, F-75205 Paris Cedex 13, France
\thanks{UMR 7164 (CNRS, Universit\'e Paris VII, CEA, Observatoire de Paris)}} \affiliation{
$^{13}$
Dublin Institute for Advanced Studies, 5 Merrion Square, Dublin 2,
Ireland} \affiliation{
$^{14}$
Landessternwarte, Universit\"at Heidelberg, K\"onigstuhl, D 69117 Heidelberg, Germany} \affiliation{
$^{15}$
Laboratoire de Physique Th\'eorique et Astroparticules, 
Universit\'e Montpellier 2, CNRS/IN2P3, CC 70, Place Eug\`ene Bataillon, F-34095
Montpellier Cedex 5, France} \affiliation{
$^{16}$
Universit\"at Erlangen-N\"urnberg, Physikalisches Institut, Erwin-Rommel-Str. 1,
D 91058 Erlangen, Germany}
% ...

\begin{abstract}

  \normalsize

  The inner 10~pc of our galaxy contains many counterpart candidates
  of the very high energy (VHE; $> 100$~GeV) \gr\ point
  source \hgcfull.  Within the point spread function of the \hess\
  measurement, at least three objects are capable of accelerating
  particles to very high energies and beyond, and of providing the observed
  \gr\ flux. Previous attempts to address this source confusion
  were hampered by the fact that the projected distances between those
  objects were of the order of the error circle radius of the emission
  centroid (34'', dominated by the pointing uncertainty of
  the \hess\ instrument). Here we present \hess\ data of the Galactic
  Centre region, recorded with an improved control of the instrument
  pointing compared to \hess\ standard pointing procedures. Stars
  observed during \gr\ observations by optical guiding cameras mounted
  on each \hess\ telescope 
  are used for off-line pointing calibration, thereby decreasing the
  systematic pointing uncertainties from 20'' to 6'' per axis. The position of
  \hgc\ is obtained by fitting a multi-Gaussian profile to the
  background-subtracted \gr\ count map. A spatial 
  comparison of the best-fit position of \hgc\ with the position and
  morphology of candidate counterparts is performed.  The position is,
  within a total error circle radius of 13'', coincident with the
  position of the supermassive black hole \astar\ and the recently
  discovered pulsar wind nebula candidate \pwn. It is significantly
  displaced from the centroid of the supernova remnant \aeast,
  excluding this object with high probability as the dominant
  source of the VHE \gr\ emission.
  \vspace{5mm}
\end{abstract}

\maketitle
%

\normalsize

\section{VHE $\gamma$-rays from the Galactic Centre}
\label{Introduction} 

Since the discovery of the strong compact radio source \astar\
\citep{Balick:1974aa}, the Galactic Centre (GC), as the closest
galactic nucleus, has served as a unique laboratory for investigating
the astrophysics of galactic nuclei in general. The radio picture
\citep{LaRosa00} of the central few 100~pc around the centre of
the Milky Way exhibits a complex and very active region, with numerous
sources of non-thermal radiation, making this region a prime target
for observations at very high energies (VHE; $> 100$~GeV).  Indeed
several Imaging Atmospheric Cherenkov Telescopes (IACTs) have 
detected a source of VHE \grs\ in the direction of the GC
\citep{Aharonian:2004wa,Albert:2005kh,Kosack:2004ri,Tsuchiya:2004wv}.
The \hess\ instrument \citep[see][and references therein]{crab}
provides the to date most precise VHE data on this source,
henceforth called \hgcfull.  As shown with deep observations in 2004,
\hgc\ is a point source for \hess\ (rms spatial extension $< 1\farcm 2$ at 95\%
CL), and is within $7''\pm 14''_{\mathrm{stat}}\pm
28''_{\mathrm{sys}}$ positionally coincident with the bright radio
source \astar\ \citep{Aharonian:2006wh}. The measured energy spectrum
does not fit Dark Matter (DM) model spectra -- at least for
the most popular models of DM annihilation --,
ruling out the bulk of the TeV emission soley to be of a DM origin
\citep{Aharonian:2006wh}. 

Of all possible astrophysics counterparts, the $3\times 10^6\
\mathrm{M}_{\astrosun}$ supermassive black hole (SMBH) coincident with the
\astar\ radio position is a compelling candidate. Various
models predict VHE emission from this object, produced either
close to the SMBH itself \citep{Aharonian:2005ti}, within an ${\cal O}(10)$~pc zone
around \astar\ due to the interaction of run-away protons with the
ambient medium \citep{Aharonian:2005b,Liu2006a,Wang:2009}, or by
electrons accelerated in termination
shocks driven by winds emerging from within a couple of Schwarzschild
radii \citep{Atoyan2004}. \astar\ is a source of bright and
frequent X-ray and infrared flares. Detection of quasi-periodic
oscillations (QPOs) on time scales of 100-2250~s has been claimed
\citep[e.g.][]{Baganoff2001,Genzel:2003,Porquet2003}. Recently,
however, observations with the Keck II telescope could not confirm the
existence of such QPOs \citep{Meyer:2008}. No hint for variability, flaring
activity, or QPOs has been found 
in the VHE \gr\ lightcurve in 93~h live time of \hess\ data collected
during the years 2004-2006 \citep{GCSpectrum}. Moreover, during a
campaign of simultaneous \hess\ and Chandra observations of \astar\ in
2005, a major X-ray flare of 1600~s duration was observed. Although the
X-ray flux increased to $\approx 9$ times the quiescent level, no
evidence for flaring activity was detected in the VHE lightcurve
\citep{Aharonian:2008yb}. This result makes it highly unlikely that
X-ray and VHE emission originate from the same source region, and puts
constraints on models predicting correlated flaring.

Besides \astar\ and its immediate vicinity, there are at least two
other production site candidates for VHE emission. The first one is
the radio-bright, shell-like supernova remnant (SNR) \aeast, which
surrounds partially \astar. SNRs have been shown to be efficient
particle accelerators \citep[see e.g.][]{Helder:2009fm}, and the
presence of an ${\cal O}$(mG) magnetic field \citep{Yusef96} makes
\aeast\ a compelling candidate for particle acceleration to very high
energies \citep{Crocker2005}. The second one is the recently detected
pulsar wind nebula (PWN) candidate \pwn\ \citep{Wang:2005ya}. Despite
of its faint X-ray flux, it may plausibly emit TeV \grs\ at an energy
flux level compatible with the \hess\ observations
\citep{Hinton:2006zk}, assuming that \pwn\ is located at the same
distance as \astar. 

A firm identification of \hgc\ is particularly hampered by the --
compared to radio or X-ray instruments -- modest angular resolution of
the current generation of Cherenkov telescopes ($\leq 5'$ for a single
\gr\ at TeV energies), which gives rise to source confusion in this
densely populated region of the galaxy. Adopting a distance to the GC of 8.33~kpc
\citep{Gillessen:2009}, the \hess\ source size upper limit encloses
a region of about 2.9~pc radius. Comparing this number to the
projected distance of \astar\ to the radio maximum of \aeast\ and the
X-ray maximum of \pwn\ (3.7~pc and 0.4~pc, respectively), it becomes
clear that a precise position measurement of the centre-of-gravity of
\hgc\ can help to shed light on the nature of this source.

Although previous \hess\ position measurements have been
unprecedentedly precise,
the relatively large -- compared to statistical errors -- 
systematic errors due to pointing uncertainties of the \hess\ array
rendered the identification of the major contributing source of the
VHE emission, and especially a clear statement on the role of \aeast,
difficult. In this paper a refined measurement of \hgc's emission
centroid is reported. Using improved telescope pointing control, the
systematic error of the measurement is decreased by a factor of three
compared to previous results, and the total error on the centroid
position is reduced to $13''$ (68\% containment radius), compared to
$34''$ in \cite{Aharonian:2006wh}. 

\section{Astrometric pointing corrections}
The 12~m mirror dishes of the \hess\ telescopes are supported by
altitude/azimuth mounts. During \gr\ observations, all four \hess\
telescopes track the targeted source with a nominal precision of
better than a few seconds of arc per axis. However, due to the weight
of the mirrors and the Cherenkov cameras, ${\cal O}$(mm) deflections of
the 15~m long camera masts and the mirror dishes make astrometric corrections
necessary. The \hess\ pointing corrections are based on the assumption
that telescope deformations, and hence pointing deviations, are
reproducible, and depend only on the (alt-az) pointing position. They
are of the order of a few minutes of arc and are applied to the
recorded events after data taking, based on a set of
independently recorded calibration data \citep{Gillessen:2004tc}.
These \emph{standard pointing corrections}, by default applied to all \hess\ data,
provide a localisation of point-like \gr\ sources with a systematic
pointing error of $20''$ per axis. 

The analysis presented here improves significantly upon these systematic
uncertainties by utilising optical guiding telescopes mounted on
the mirror dish of each \hess\ telescope. 
Stars within the field of view ($0.3^\circ\times 0.5^\circ$)
of the guiding telescopes are imaged by CCD cameras with a projected pixel size
of $2\farcs3$.  The guiding telescope optics is slightly defocused such that the
light from each star is imaged onto several CCD pixels, making precise
position measurement possible. Positions of recorded stars are then compared to
nominal coordinates listed in the Hipparcos and Tycho star catalogues \citep{HIP:1997}.
For the analysis presented here, images were recorded at a rate of
about 1/min and contain typically 2-10 identified stars.
Additionally, for each \hess\ telescope, deformations of the
Cherenkov camera masts are measured by monitoring eight reference LEDs
mounted on the Cherenkov camera body. This is done with the help of
CCD cameras installed at the centre of each mirror dish.
From the combined information of the two CCD
cameras, pointing corrections are calculated for the individual \hess\
telescopes. To correct the direction of
each \gr, linear interpolation is used between the pointing
corrections derived from the individual CCD images. The difference in
refraction correction for star light and Cherenkov light from \gr\
showers is taken into account.
An absolute calibration of the guiding system is performed 
at the beginning and end of every moon period: the telescopes are
pointed at typically 50 bright stars uniformly distributed in the sky.
Images of the stars are recorded with the guiding telescopes.
Additionally, the star light is reflected by the mirrors onto screens
in front of the Cherenkov cameras, and images of the stars and of the
reference LEDs are recorded with the central CCD cameras. From these
measurements altitude and azimuth dependent pointing models are
derived. These relate, for any given observation position, the star
position measured with the guiding telescopes to the star spot
position determined with the central CCD camera. 

Typically, the \emph{precision pointing corrections} derived in this
way differ only slightly from the standard pointing corrections.
However, the observation of stars and camera body simultaneously to \gr\
collection reduces systematic uncertainties significantly, such as
hysteresis effects observed in the camera mast structure, which limit
the precision of the altitude determination. Systematic errors due to
an observed long-term movement of the telescope foundations are
cancelled, as are uncertainties in the absolute positioning of the
tracking system. Thermal expansion of the CCD chips due to
changes of the ambient temperature is accounted for in the precision
pointing model. For the data set presented in this analysis, a total
systematic pointing error of $\pm 6''$ per axis on the sky
was derived \citep{braun07}. Possible
systematic effects regarding the reconstruction of the \gr\ shower
images, such as an inhomogeneous field-of-view or the effect of
Earth's magnetic field on the image parameters have been studied. No
effect was observed that would systematically shift the centroid of
point-like \gr\ sources by more than $2''$. Furthermore, the precision
pointing corrections were extensively validated on VHE $\gamma$-ray
data of point-like sources with positions and extensions known from observations
at another waveband (with much better pointing accuracy and angular
resolution). A detailed description of the
precision pointing technique and the estimation of systematic errors
is beyond the scope of this letter and will be published elsewhere.

\begin{figure*}
  \includegraphics[width=\textwidth]{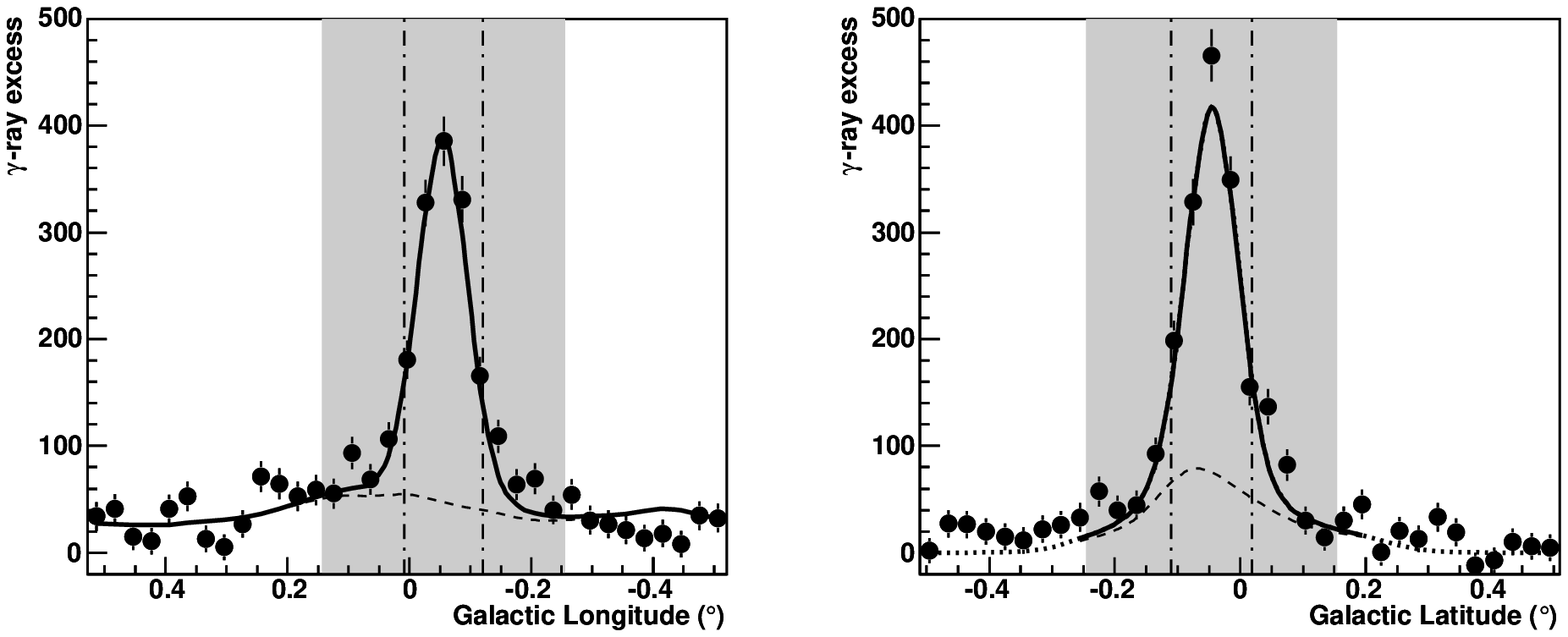}
  \caption{Projections along Galactic longitude (left) and
    Galactic latitude (right) of the acceptance corrected \gr\ excess
    map and the best-fit function (solid line). The
    latitude and longitude ranges used 
    for the projections correspond to the respective fit ranges and are
    indicated by the shaded regions. The 68\% containment region of
    the PSF of the \hess\ instrument
    is shown by the dashed-dotted lines.
    The fitted contribution of diffuse \gr\ emission is indicated
    by the dashed curve. The dotted lines in the latitude projection
    depict the extension of the best-fit function beyond
    the latitude coverage of the CS line emission survey used for the
    diffuse \gr\ 
    background model. Possible features in the \gr\ excess outside of
    the fit boundaries are under study, but do not affect the position
    determination.  
}
\label{fig:FitSlices}
\end{figure*}

\section{Analysis of $\gamma$-ray data}

Since the guiding telescopes for precision pointing corrections are in
operation only since 2005, the results reported here are based on 64~h
(live time) of data recorded with the \hess\ instrument between May 4
and August 23, 2005, and between April 4 and August 4, 2006. Most of
the observations (59~h) were carried out in \emph{wobble mode}, i.e.
the telescope pointing direction was offset from the target direction
(\astar) by typically $0.5^\circ-0.7^\circ$ in either right ascension or
declination, in an alternating fashion.  The remaining 5~h of data were
recorded with various offsets of up to $1.4^\circ$ from the direction
of \astar.  The mean zenith angle of the data described here is
$23^\circ$, and the observation zenith angles range from
$6^\circ-60^\circ$.

Data were analysed with the standard \hess\ calibration and
reconstruction chain \citep{crab}. First, each shower image recorded
by the Cherenkov cameras was corrected for astrometry using the
precision pointing corrections described above. To suppress background
events caused by cosmic ray induced air showers, \grs\ were selected
based on the shape of the shower images in the Cherenkov cameras, as
described by Hillas parameters \citep{hillas} using \emph{hard cuts}
\citep{crab}. As opposed to standard cuts, hard cuts select
high-intensity shower images, reducing further 
the number of background events (relative to signal) at the expense of
a higher energy threshold ($\approx 630$~GeV for a mean zenith angle of
$23^\circ$). In addition, this high-intensity selection  
leads to a sample of well reconstructed showers, resulting
in an improved angular resolution. After this 
event selection, the direction of each \gr\ was reconstructed by
intersecting the major axes of the Hillas ellipses, following
algorithm 3 from \citet{hofmann}. This approach uses the Hillas width
and length of the shower images to estimate the \gr\ direction
independently with each telescope. These estimates are then combined to
yield the optimum \gr\ direction, which improves upon the standard
\hess\ reconstruction in terms of angular resolution.  Reconstructed
events were accumulated in a $2^\circ\times 2^\circ$ image, centred at
the position of \astar\ and binned into squares of $0.03^\circ$ angular
size. Remaining background from cosmic-ray induced showers was
estimated using the Ring-Background technique \citep{Berge:2006ae},
excluding regions containing known \gr\ sources, such as the band of
diffuse emission along the Galactic Centre ridge
\citep{Aharonian:2006au}.  A background subtraction based on a
template approach \citep{Rowell:2003jb} gives consistent results. An
excess of $1313\pm 42$ VHE \grs\ is found within a circle of
radius $0.1^\circ$ centred on \astar, with a statistical significance
of 46 standard deviations above the background. The energy spectrum
derived from this reduced data set is compatible
with that reported in \citet{GCSpectrum}, which was obtained with a
different analysis chain.

The point spread function (PSF), reflecting the angular extension of a
point source seen by the \hess\ instrument, was modeled using
Monte-Carlo \gr\ simulations, taking into account the distributions of
zenith angle, of offset of the pointing position relative to
\astar, as well as the energy distribution of \grs\ from \hgc\
\citep{GCSpectrum}. The simulated PSF can be well described by the
sum of two Gaussian functions with equal mean \citep{crab}. The
overall angular resolution of the data set is $3\farcm 9$ (68\%
containment radius). 

\section{Position of \hgcfull}

The centroid of the VHE emission was determined by fitting the
acceptance corrected and background subtracted \gr\ count map in a
window of $\pm 0.2^\circ$ centred on \astar, with a 2-dimensional
radially symmetric profile. The fit model was composed of a
double-Gaussian part accounting for
the PSF of the \hess\ instrument, convolved with an assumed Gaussian
surface brightness distribution to account for a possible intrinsic
extension of the source. This source extension and the overall
normalisation were left as free parameters in the fit. The PSF was
fixed from MC simulations as described above.

Diffuse \gr\ emission along the Galactic Centre ridge introduces an
asymmetric \gr\ background in the region of \astar, which could in
principle bias the position determination of \hgc.  In a circular region of
$0.1^\circ$ around the centroid position, this background is at a
level of $\approx 15$\% of the total flux observed from the source
\citep{Aharonian:2006wh,GCSpectrum}.  In the position determination,
the background \gr\ diffuse emission was therefore taken into account
by adding an independent term in the fit function with free
normalisation. The expected diffuse \gr\ emission was modelled --
following \citet{Aharonian:2006au} -- by a radially symmetric
Gaussian distribution of
width $0.8^\circ$ centred at the GC, multiplied with the density
distribution of molecular clouds in the region from CS line emission
measurements \citep{Tsuboi:1999a}. The position fit is largely
insensitive to details of the diffuse emission model.  Indeed, when
fitting the position of \hgc\ without taking into account the diffuse
component, the result is still consistent within statistical errors
with the final position quoted below.

Using a $\chi^2$-minimisation procedure, the best-fit position of \hgc\ in
equatorial coordinates is \position\ (J2000.0). The best-fit
probability is 12\%. The best-fit position is within
\distToAStar\ coincident with the position of \astar, and fully compatible with
the position reported from the 2004 data set \citep{Aharonian:2006wh}.
Changing the background subtraction technique, the image binning, or the fit
boundaries did not affect the position by more than $2''$. Assuming a
Gaussian distribution of surface brightness, a rms source size upper
limit of $1\farcm 3$ (95\% CL) is 
derived.  The fraction of diffuse emission in a circle of $0.1^\circ$
radius centred on the best-fit position is 14\%, consistent with
previous findings \citep{Aharonian:2006wh,GCSpectrum}. Fig.~
\ref{fig:FitSlices} shows longitude and latitude slices of the \gr\
excess map, with the best-fit function overlaid, to demonstrate
the performance of the fit.

\section{Discussion}

\begin{figure}
  \includegraphics[width=0.49\textwidth]{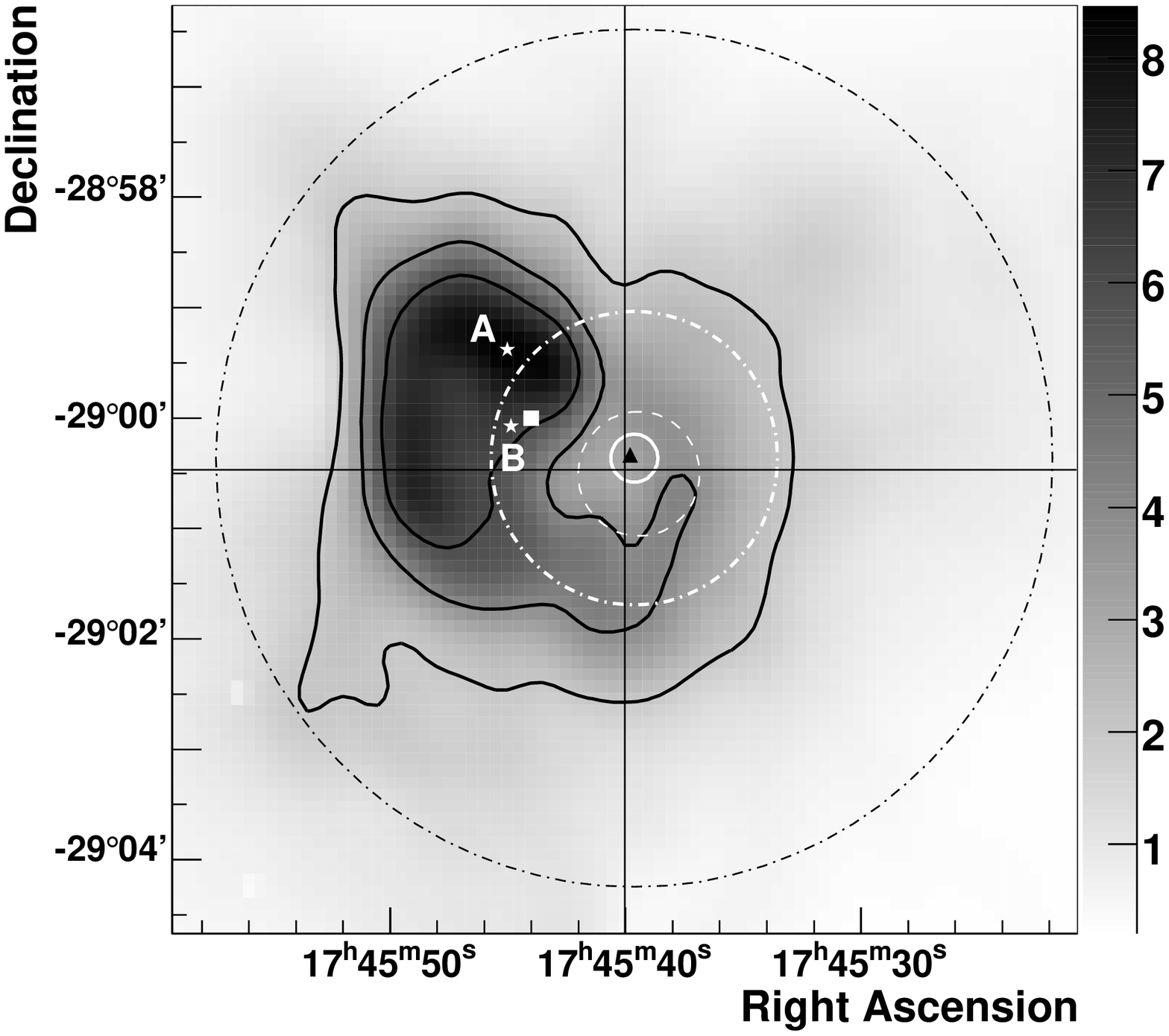}
  \caption{90~cm VLA radio flux density map \citep{LaRosa00} of the
    innermost 20~pc of the GC, showing emission from the SNR \aeast.
    Black contours denote radio flux levels of 2, 4, and 6~Jy/beam.
    The centre of the SNR \citep{Green:2009qf} is marked by the white
    square, and the positions of \astar\ \citep{saga_radio} and \pwn\
    \citep[head position,][]{Wang:2005ya} are given by the cross hairs
    and the black triangle, respectively. The 68\% CL total error contour of
    the best-fit centroid position of \hgc\ is given by the white
    circle. The dashed white circle shows the same contour for the
    previously reported \hess\ measurement \citep{Aharonian:2006wh}.
    The white and black dashed-dotted lines show the 95\% CL upper limit
    contour of the source extension and the 68\% containment region of
    the \hess\ PSF, respectively. The white stars marked \emph{A} and
    \emph{B} denote the position of the radio maximum and the best-fit
    position for the radio emission after smoothing with the PSF of the
    \hess\ instrument, respectively.  }
\label{fig:RadioMap}
\end{figure}

Fig.~\ref{fig:RadioMap} shows a VLA 90~cm image of the innermost 20~pc
region of the GC, centred on \astar. The shell-like radio structure of the SNR \aeast\ is
clearly visible. The best-fit position of \hgc\ lies in a region where
the radio emission is comparatively low, and is shown as a 68\%~CL
total error contour, computed from the summed (in quadrature)
statistical and systematic best-fit position errors. As can be seen
from the figure, the centroid of the VHE source is 
coincident with the positions of \astar\ and \pwn, but inconsistent
with the regions of intense radio emission from \aeast. Two rather
independent approaches to derive a quantitative statement about the
compatibility of \hgc's best-fit position with \aeast\ are detailed in the
following.

The white star labelled $A$ in Fig.~\ref{fig:RadioMap} denotes the
position of \aeast's radio maximum. Comparing the 68\%~CL radius of the
observed VHE centroid to the angular distance between $A$ and the
best-fit position, a coincidence of the two positions is ruled out
with 7.1 standard deviations. By the same
arguments, VHE point emission from the centre of the SNR
\citep{Green:2009qf}, indicated as a white square in
Fig.~\ref{fig:RadioMap}, is ruled out with 4.7 standard deviations.

Instead of point emission, extended emission can be
considered, e.g. by assuming that the hypothetical VHE emission from
\aeast\ follows closely the morphology of the radio flux. The centroid
of such emission would be detected at the coordinates marked $B$ in
Fig.~\ref{fig:RadioMap}. This position was derived by fitting the
radio map -- smoothed with the \hess\ PSF -- with the technique used
above for the VHE \gr\ data. Following the methods used for position
$A$, the radio fit position is 5.4 standard deviations away from the
best-fit VHE centroid position.

The above results are obtained assuming that the VHE emission and
radio morphology are correlated. Since the best-fit position
does not coincide with a region of intense radio emission (see Fig.~
\ref{fig:RadioMap}), relaxing this assumption leads to more
conservative estimates of the association probability.  A priori, it
would appear conservative to assume that the centroid of TeV emission
associated with \aeast\ might appear anywhere within the boundaries,
with equal probability.
With this assumption one can calculate the probability that the VHE
emission is produced inside \aeast, but is only by chance positionally
coincident with \astar\ and \pwn, which themselves are plausible
emitters of VHE radiation and thus viable counterpart candidates of
\hgc. Defining the 2~Jy/beam radio contour of \aeast\ as the SNR
boundary, which encloses the best-fit position of the emission
centroid, a chance probability of $9\times 10^{-5}$ is derived
(corresponding to 3.9 standard deviations). This
number does slightly change depending on the assumed size of the SNR
boundary.  It is clear, however, that even with this conservative
approach an association of \aeast\ with the observed VHE \gr\ emission
is rather unlikely.

The exclusion of \aeast\ as the main contributor to the VHE emission
is a major step towards an identification of \hgc.  Despite the fact that
\hgc\ is a non-variable \gr\ source, both \pwn\ and \astar\ are compelling
counterpart candidates, as models
exist (see section \ref{Introduction}), which can explain a steady
$\gamma$-ray flux and variable X-ray emission from \astar\ at the same
time.  More information is needed to discriminate between these two
objects. Due to their enhanced angular resolution and sensitivity, and
their extended energy range, proposed future 
VHE \gr\ observatories such as CTA or AGIS could shed light on the
open question of which of these sources dominates the production of
the VHE \gr\ emission from the gravitational centre of our galaxy.

\section*{Acknowledgements}
  The support of the Namibian authorities and of the University of
  Namibia in facilitating the construction and operation of H.E.S.S.
  is gratefully acknowledged, as is the support by the German Ministry
  for Education and Research (BMBF), the Max Planck Society, the
  French Ministry for Research, the CNRS-IN2P3 and the Astroparticle
  Interdisciplinary Programme of the CNRS, the U.K. Science and
  Technology Facilities Council (STFC), the IPNP of the Charles
  University, the Polish Ministry of Science and Higher Education, the
  South African Department of Science and Technology and National
  Research Foundation, and by the University of Namibia. We appreciate
  the excellent work of the technical support staff in Berlin, Durham,
  Hamburg, Heidelberg, Palaiseau, Paris, Saclay, and in Namibia in the
  construction and operation of the equipment.  

\bibliographystyle{mn2e}
\bibliography{GCPositionPaper}

\end{document}